\documentclass{emulateapj}
\usepackage{amssymb}
\usepackage{apjfonts} 
\usepackage{natbib}

\newcommand{\htn}{H$^{13}$CN}
\newcommand{\ratio}{H$^{12}$CN/H$^{13}$CN}
\newcommand{\ratioc}{$^{12}$C/$^{13}$C}
\newcommand{\be}{\begin{equation}}
\newcommand{\ee}{\end{equation}}

\usepackage[utf8]{inputenc}
\usepackage{rotating}
\usepackage{multirow}
\usepackage{pinlabel} 

\shorttitle{ALMA Autocorrelation Spectroscopy of Comets}
\shortauthors{Cordiner et al.}

\begin{document}

\title{ALMA Autocorrelation Spectroscopy of Comets: \\The HCN/H$^{13}$CN ratio in C/2012 S1 (ISON)}

\author{M. A. Cordiner\altaffilmark{1,2}, M. Y. Palmer\altaffilmark{1,2}, M. de Val-Borro\altaffilmark{1,2}, S. B. Charnley\altaffilmark{1}, L. Paganini\altaffilmark{1,2}, G. Villanueva\altaffilmark{1}, D. Bockel{\'e}e-Morvan\altaffilmark{3}, N. Biver\altaffilmark{3},  A. J. Remijan\altaffilmark{6}, Y.-J. Kuan\altaffilmark{4,5}, S. N. Milam\altaffilmark{1}, J. Crovisier\altaffilmark{3}, D. C. Lis\altaffilmark{7}, M. J. Mumma\altaffilmark{1}}

\altaffiltext{1}{NASA Goddard Space Flight Center, 8800 Greenbelt Road, Greenbelt, MD 20771, USA.}
\altaffiltext{2}{Department of Physics, Catholic University of America, Washington, DC 20064, USA.}
\altaffiltext{3}{LESIA, Observatoire de Paris, PSL Research University, CNRS, Sorbonne
Universit, Univ.  Paris-Diderot, Sorbonne Paris Cit{\'e}, 5 place Jules Janssen, 92195 Meudon, France}
\altaffiltext{4}{National Taiwan Normal University, Taipei 116, Taiwan, ROC.}
\altaffiltext{5}{Institute of Astronomy and Astrophysics, Academia Sinica, Taipei 106, Taiwan, ROC.}
\altaffiltext{6}{National Radio Astronomy Observatory, Charlottesville, VA 22903, USA.}
\altaffiltext{7}{Sorbonne Universit{\'e}, Observatoire de Paris, Universit{\'e} PSL, CNRS, LERMA, F-75014, Paris, France.}

\begin{abstract}
The Atacama Large Millimeter/submillimeter Array (ALMA) is a powerful tool for high-resolution mapping of comets, but the main interferometer (comprised of 50$\times12$-m antennas) is insensitive to the largest coma scales due to a lack of very short baselines.  In this work, we present a new technique employing ALMA autocorrelation data (obtained simultaneously with the interferometric observations), effectively treating the entire 12-m array as a collection of single-dish telescopes. Using combined autocorrelation spectra from 28 active antennas, we recovered extended HCN coma emission from comet C/2012 S1 (ISON), resulting in a fourteen-fold increase in detected line brightness compared with the interferometer. This resulted in the first detection of rotational emission from H$^{13}$CN in this comet. Using a detailed coma radiative transfer model accounting for optical depth and non-LTE excitation effects, we obtained an H$^{12}$CN/H$^{13}$CN ratio of 88$\pm$18, which matches the terrestrial value of 89, consistent with a lack of isotopic fractionation in HCN during comet formation in the protosolar accretion disk. {The possibility of future discoveries in extended sources using autocorrelation spectroscopy from the main ALMA array is thus demonstrated.}
\end{abstract}


\section{Introduction}
The Oort Cloud is a reservoir of comets sufficiently distant from the sun to have experienced little thermal alteration since the formation of the Solar System \citep{gui15}. Studies of Oort-Cloud cometary materials can therefore provide unique information on the composition of the protosolar disk, from which insights into the physical and chemical processes occurring during planet formation may be obtained.  
    
Molecular isotopologue ratios encode information on the thermal and chemical histories of a wide variety of solar system materials \citep[\emph{e.g.}][]{boc15,mar18,ale18}, and provide fundamental insights into their origins. In the cold interstellar medium, isotopic fractionation occurs as a result of gas-phase kinetic isotope effects \citep{lan84,mil89}. The difference in zero point energy between the reactants and products means that at low temperatures (below the activation energy for the reverse reaction), isotopic exchange reactions such as 

\be
^{13}{\rm C}^+ + ^{12}{\rm CO} \leftrightharpoons ^{12}{\rm C}^+ + ^{13}{\rm CO} 
\ee

\noindent tend to proceed preferentially in the forward direction \citep[see][]{rou15}. Strong isotopic fractionation effects are observed in species containing carbon \citep{sak13}, deuterium \citep{cec14}, and nitrogen \citep{hil13} in the local interstellar medium (ISM).   Consequently, during the early history of the Solar System (and prior interstellar phase), it may be expected that the gases eventually incorporated into icy planetesimals and comets were enriched (or depleted) in heavy isotopes. 

Isotopic fractionation is routinely observed for nitrogen-bearing cometary species. The average $^{14}$N/$^{15}$N ratio in cometary NH$_2$ was found to be 136 \citep{shi16}, with values $\sim130-170$ for CN and HCN \citep{boc15}. These values are significantly smaller than both the terrestrial and protosolar values of 272 and 440, respectively \citep{mar12}, highlighting the likely importance of low-temperature processing, resulting in $^{15}$N-enrichment during the formation of these nitrogen-bearing compounds.
    
So far, measurements of cometary $^{12}$C/$^{13}$C ratios have provided little evidence for fractionation. The ratios of $^{12}$CN/$^{13}$CN in 23 comets,$^{12}$C$_2$/$^{13}$C$^{12}$C in 7 comets \citep{boc15}, and CO$_2$ in comet 67P \citep{has17} are consistent with the terrestrial $^{12}$C/$^{13}$C value of 89. By contrast, the H$^{12}$CN/H$^{13}$CN ratio has been measured in only a few comets to-date \citep[\emph{e.g.}][]{jew97,boc08,biv16}. This is primarily due to the difficulty in reliably detecting the weak emission lines from H$^{13}$CN. Using the most reliable data on three comets, \citet{biv16} concluded that the $^{12}$C/$^{13}$C ratio in HCN may be somewhat larger than terrestrial, but additional high-sensitivity observations are required to confirm this possibility. The need for improved constraints on the $^{13}$C fraction in cometary HCN provides motivation for our present study.

We also seek to demonstrate the first scientific use of autocorrelation observations from the Atacama Large Millimeter/sub-mm Array (ALMA) main array. {The main ALMA array comprises $50\times12$ m antennas used for interferometry. There are an additional $12\times7$ m antennas in the Atacama Compacy Array (ACA), and four 12 m antennas for total power (autocorrelation) observations. Combining all 66 antennas can provide images that are complete on all spatial scales. However, the total-power antennas were not yet commissioned at the time of our 2013 observations of comet C/2012 S1 (ISON).}  

The main ALMA array (used in our present study), in interferometric mode, is insensitive to emission from structures with angular sizes greater than $\lambda/d$, where $\lambda$ is the wavelength and $d$ is the distance between the closest pair of antennas.  During typical cometary apparitions (at distances $\lesssim1$~au from Earth), cometary molecular comae may extend to sizes more than several hundred arcseconds. Consequently, with a maximum recoverable angular scale of $\lesssim10''$ in the sub-millimeter band, ALMA interferometric observations are blind to the majority of this emission. Here, we present a new technique utilizing autocorrelation (total power) data from all 28 antennas active {in the main array} during our ALMA observations of comet ISON, to recover extended flux lost by the interferometer. This resulted in a dramatic improvement in sensitivity to weak spectral lines, leading to the first detection of H$^{13}$CN in this comet, and allowing us to derive the H$^{12}$CN/H$^{13}$CN production rate ratio.

\section{Observations and Calibration}
Comet C/2012 S1 (ISON) is a dynamically new Oort Cloud comet that entered the inner Solar System for the first time in 2013. We obtained ALMA observations using the Band 7 receiver on 2013 November 17 at a heliocentric distance of 0.54 au and geocentric distance 0.88 au (see \citealt{cor14,cor17} for further details).  The observing sequence consisted of a series of $\approx7$ min scans of the comet, interleaved with $\approx1$ min scans of the phase-calibration quasar 3C\, 279, which we later used as an improvised off-source position for sky subtraction. The correlator was configured with two frequency settings (centered around 350 and 357 GHz), observed during UT 11:31-12:16 and UT 12:30-13:29, respectively. Setting 1 covered HCN and H$^{13}$CN, while setting 2 covered three CH$_3$OH lines (see Table \ref{tab:lines}).  Weather conditions were excellent for all observations, with high atmospheric stability and low precipitable water vapor (PWV = 0.52-0.57 mm at zenith). The median system temperature was $T_{sys}=134$~K.  Twenty-eight 12-m antennas were active and the minimum antenna spacing was 17.3~m, which corresponds to a maximum recoverable angular scale of $\sim$5" in the interferometric images. The primary beam FWHM was $16.4''$ at 354~GHz.

During conventional position-switched single-dish radio observations, an emission-free reference position is selected, close to the position of the science target, and (assuming identical sky conditions), the difference between the on and off-source observations gives the intrinsic source signal. Here we employ an ad-hoc position switching strategy utilizing the standard phase calibrator scans as the \emph{off-source} position for removal of background contamination from the ISON autocorrelation spectra. For more details regarding single-dish calibration techniques, see \citet{one02}. Although autocorrelation spectra are obtained for every antenna as part of normal ALMA interferometric observations, they are usually flagged (discarded), and consequently overlooked by interferometric observers.

The raw ALMA data were initially subject to standard calibration in CASA \citep{jae08}, using the scripts supplied by the Joint ALMA Observatory (JAO). The autocorrelation spectra for each pair of \emph{on} (source) and \emph{off} (reference) scans were unflagged, then extracted using {\tt plotms}. Background-subtracted spectra were produced for each individual antenna by taking the quotient: \textit{(on-off)/off}, then converted to antenna temperatures by multiplying by $T_{sys}$ at the frequency of interest. A correction for atmospheric opacity of $e^{\tau_0}/sin(\phi)$ was applied, where $\phi$ is the elevation angle and $\tau_0$ the zenith opacity, thus placing the spectra on the $T_A'$ scale.  Any remaining baseline ripples and continuum offsets were subsequently subtracted using a Savitsky-Golay filter to the line-free parts of the spectrum, using a 51 channel window and polynomial of order 3.

\begin{figure}
\centering
\hspace{-5.5mm}
\labellist
\pinlabel a) at 10 260
\endlabellist
\hspace{1mm}
\includegraphics[width=0.95\columnwidth]{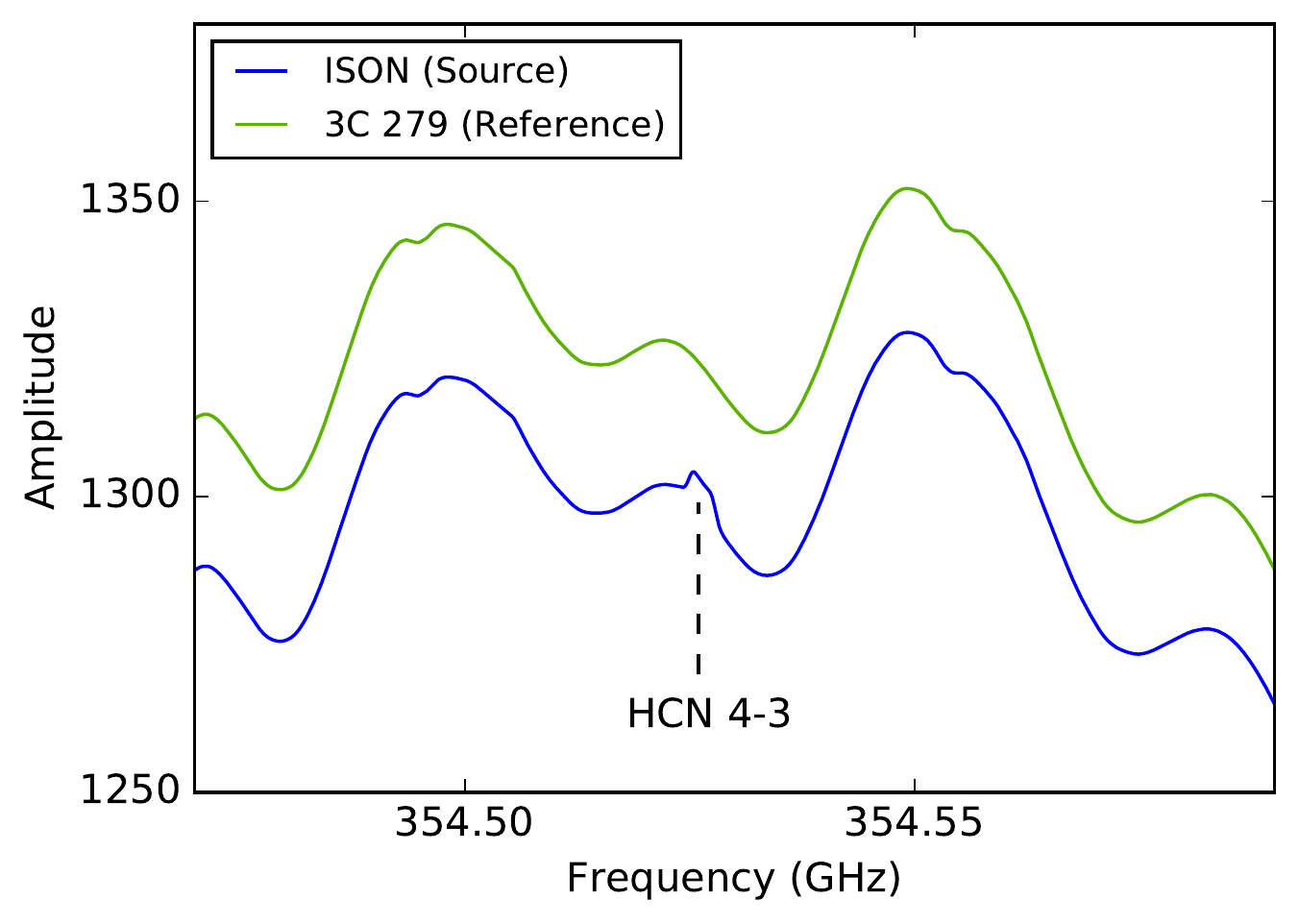}
\labellist
\pinlabel b) at 10 490
\pinlabel c) at 10 247
\endlabellist
\includegraphics[width=\columnwidth]{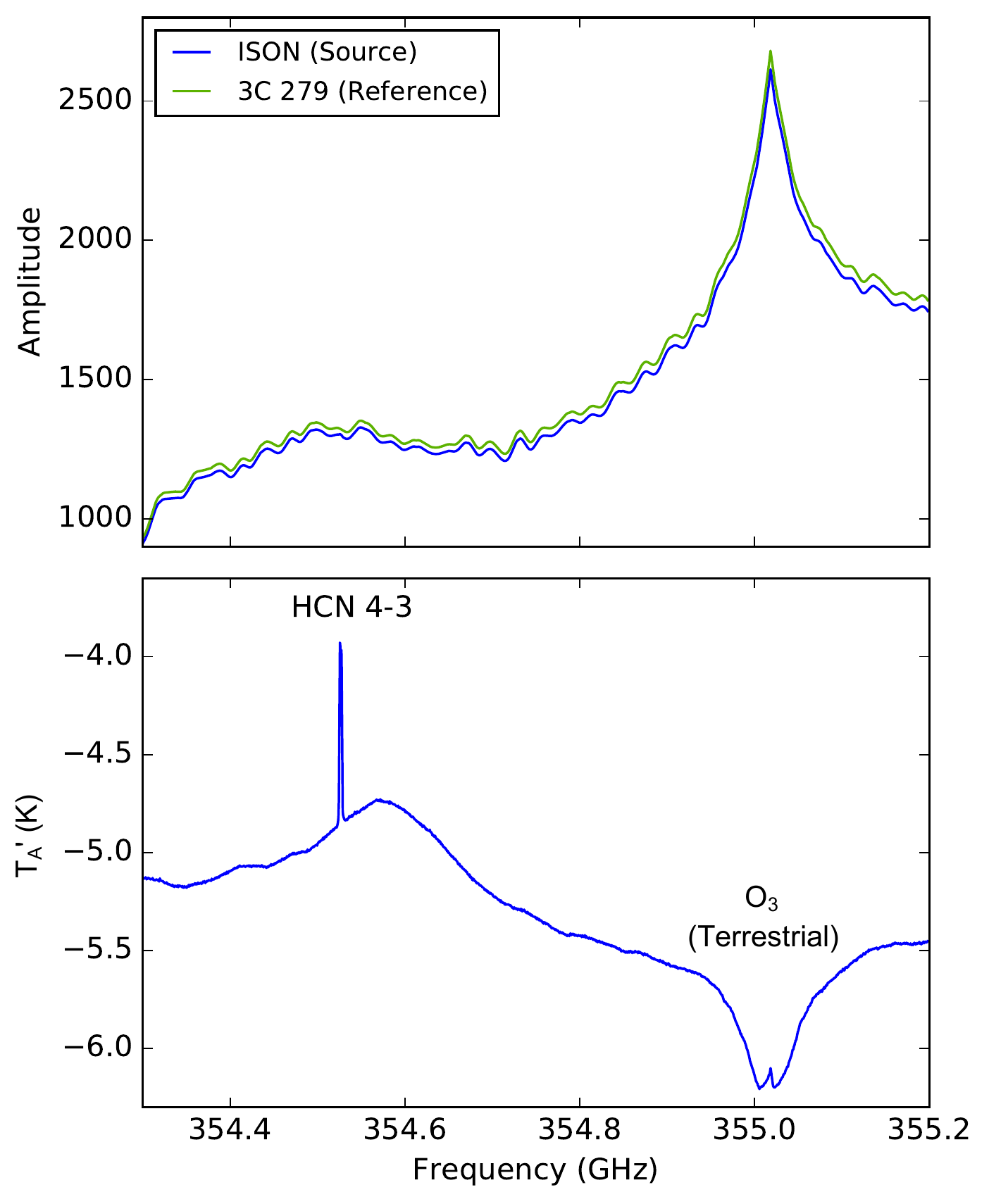} 
\caption{(a) Raw autocorrelation spectra of comet ISON (source) and quasar 3C\, 279 (reference), zoomed in on the region of the HCN ($J=4-3$) line. (b) Same as (a), but showing the full spectral window width.  (c)~Position-switched (calibrated) spectrum of comet ISON, using 3C\, 279 as the sky reference position. Cometary HCN is apparent, as well as artifacts due to incomplete sky cancellation.
\label{fig:method}}
\end{figure}

\section{Results}

Figures \ref{fig:method}a and \ref{fig:method}b show the raw autocorrelation spectra for the comet and sky-reference position (3C\, 279), averaged over all scans and all antennas.  The resulting amplitude spectra are dominated by sky emission features (such as the bright terrestrial ozone (O$_3$) line around 355 GHz), as well as ripples caused by interference between standing waves in the telescope optics, which swamp the faint signal from the comet. Figure \ref{fig:method}c shows the background-subtracted spectrum, prior to baseline removal.

Because the quasar (3C\, 279) in the sky reference field has essentially a flat spectrum over the bandwidth of our spectral regions of interest, its presence does not hinder the identification and measurement of emission lines from the comet. Furthermore, the 7~Jy flux from 3C\, 279 represents only a small fraction ($<1$\%) of the total amplitude, which is dominated by thermal emissions from the sky, telescope and receiver, so its impact on the calibrated comet spectrum is minor. Variations in sky background between the rather long (7~min) intervals between source and reference scans are expected to constitute a larger source of error. Some differences in the sky signals {are also} expected due to the relatively large ($7^{\circ}$) distance between source and reference positions. The resulting $\approx2$\% difference in airmass introduces a corresponding error in the calibrated intensity scale, and is a likely explanation for the overall difference in amplitude between the source and reference spectra (Figure \ref{fig:method}a), as well as the incomplete background removal (Figure \ref{fig:method}c).

\begin{figure}
\centering
\labellist
\pinlabel a) at 10 505
\pinlabel b) at 10 260
\endlabellist
\includegraphics[width=\columnwidth]{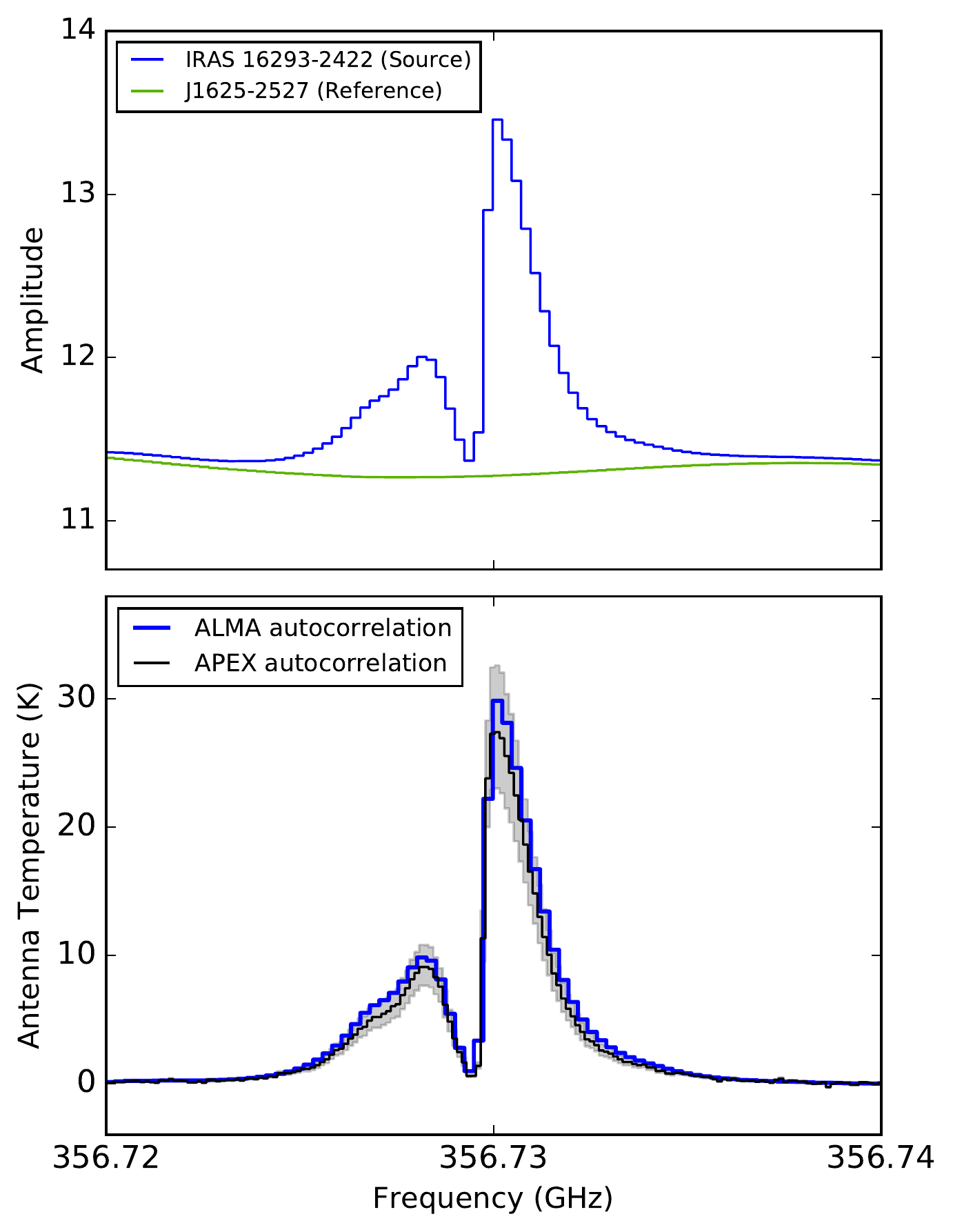}
\caption{(a) Raw ALMA autocorrelation spectra of IRAS~16293-2422 and reference quasar J1625-2527. (b) Calibrated ALMA autocorrelation spectrum of IRAS~16293-2422 HCO$^+$, compared with APEX observations. Grey shaded region is the $\pm1\sigma$ error on the APEX spectrum.
\label{fig:apex}}
\end{figure}

The reliability of our calibration method was investigated by comparing an ALMA autocorrelation spectrum to single-dish data obtained using the Atacama Pathfinder EXperiment (APEX), which has a 12-m antenna similar to that of ALMA.  We obtained observations of the protostar IRAS~16293-2422 from the ALMA science archive {(project \#2013.1.00278.S)}, covering a strong HCO$^+$ $J=4-3$ emission line, and generated the mean autocorrelation spectrum from the 12-m array as explained above. {The raw, uncalibrated spectra of IRAS~16293-2422 and reference quasar J1625-2527 are shown in Figure \ref{fig:apex}a}. IRAS~16293-2422 is routinely observed to monitor the APEX amplitude calibration, and we averaged 13 archival HCO$^+$ spectra (observed 2008 to 2017) to produce the comparison spectrum shown in Figure \ref{fig:apex}b.  The APEX and ALMA data are in excellent agreement --- our ALMA autocorrelation spectrum lies within one standard deviation of the APEX spectra (shaded region), { confirming the efficacy of our method}. 
    
\begin{figure}
\labellist
\pinlabel a) at 10 390
\endlabellist
\centering
\hspace{-1.4mm}
\includegraphics[width=0.81\columnwidth]{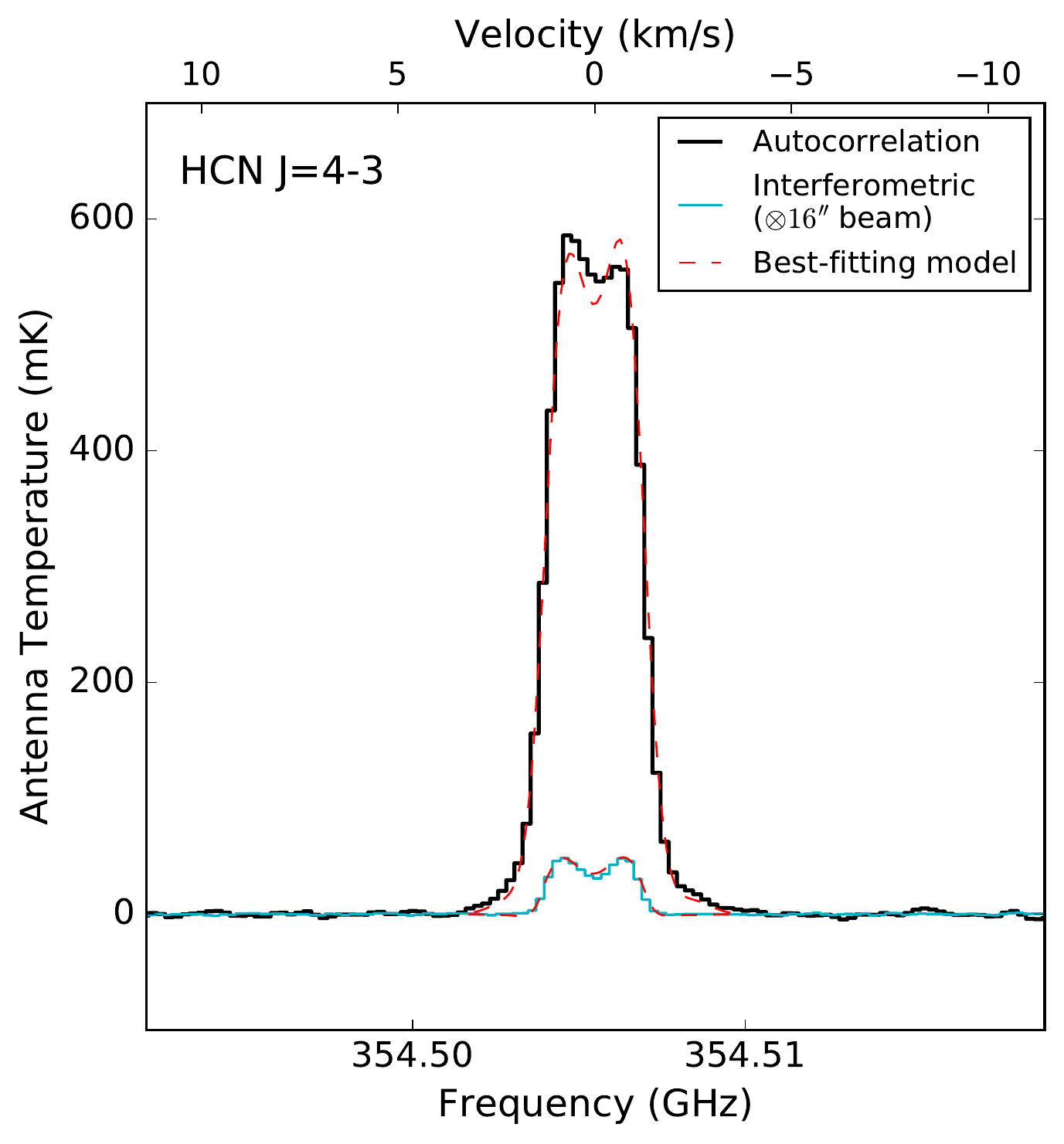}
\labellist
\pinlabel b) at 10 390
\endlabellist
\includegraphics[width=0.8\columnwidth]{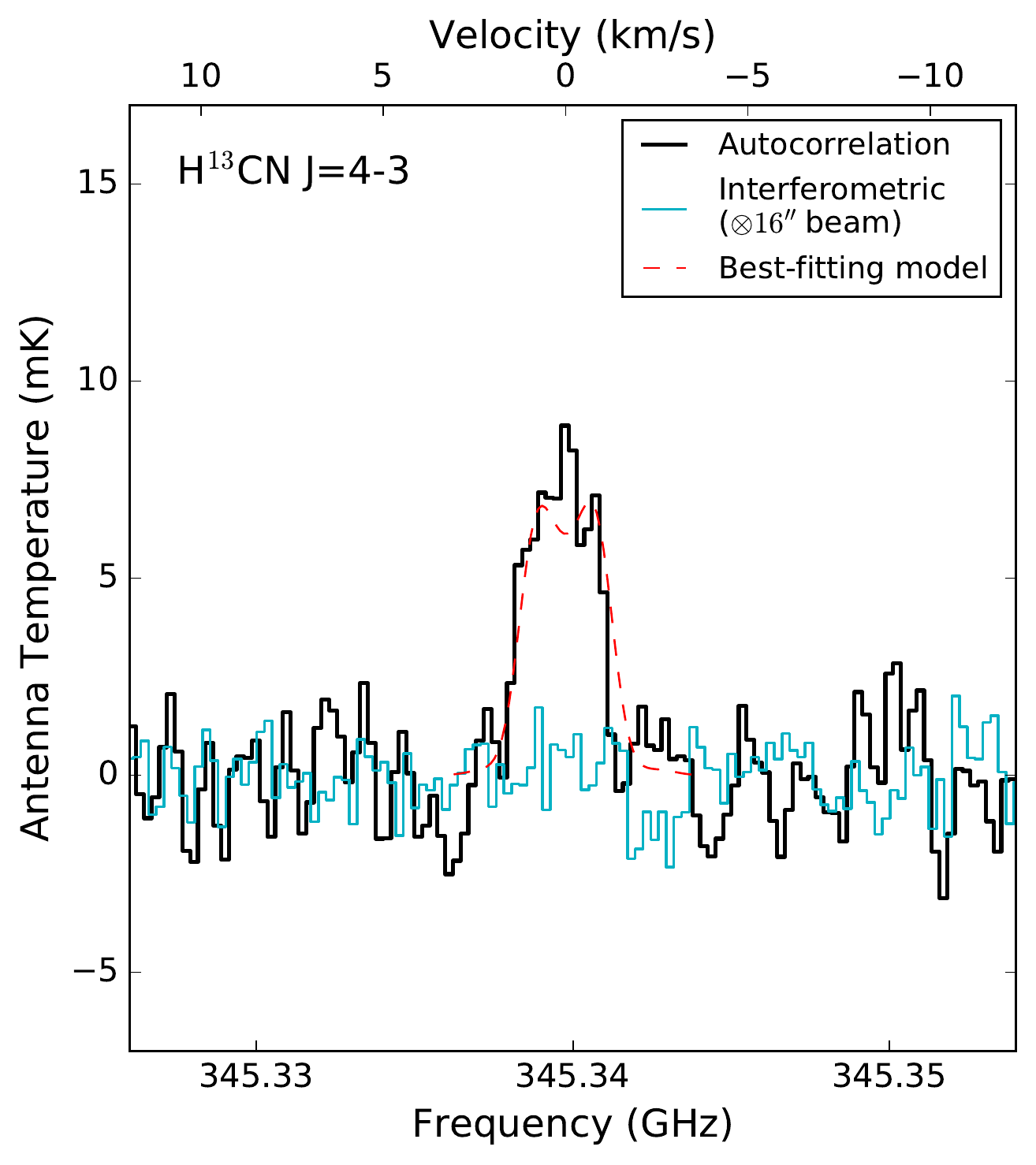}
\caption{Comparison of ALMA autocorrelation (black) and interferometric (blue) spectra for (a) HCN and (b) H$^{13}$CN. Best-fitting model spectra are overlaid (dashed red). Interferometric data were averaged over a 16.4" Gaussian aperture to match the autocorrelation beam; their intensity is greatly reduced due to {interferometric filtering}.  
\label{fig:comparison}}
\end{figure}

Baseline-subtracted HCN and \htn\ spectra {of comet ISON} are shown in Figure \ref{fig:comparison}. Interferometric (cross-correlation) spectra, averaged over the ALMA primary beam, are also shown for comparison, demonstrating a factor of 14 increase in detected line brightness using our autocorrelation technique. Note that \htn\ was not even detectable in the cross-correlation data, so the benefit of (combined) autocorrelation observations for measuring weak cometary lines is clear. The detected lines and their integrated intensities are shown in Table \ref{tab:lines}. Error estimates include the statistical noise contribution, $2$\% sky background uncertainty and an error of $\sigma\times$ FWHM (where $\sigma$ is the RMS noise), to allow for uncertainty in the baseline {fitting}. 


A line of SO$_2$ (at 345.339~GHz) has the potential to overlap and contaminate the \htn\ $J=4-3$  emission \citep{lis97}. However, the SO$_2$ 351.874~GHz and 355.046~GHz lines were in our observed bandpass, and are expected to be of similar strength to the 345.339~GHz line. We found no sign of any emission at those frequencies, so the impact of SO$_2$ on our \htn\ measurement should be negligible.

\begin{table*}
\centering
\caption{Detected Spectral Lines and Molecular Production Rates\label{tab:lines}}
{\footnotesize
\begin{tabular}{lcr@{$-$}lcccc}
\hline
Species&Setting&\multicolumn{2}{c}{Transition}&Frequency (GHz)&$E_u$ (K)&$\int{T_A'}dv$ (mK km s$^{-1}$) & Q (s$^{-1}$) \\
\hline
H$^{13}$CN&1&4&3&345.340&41.44&$18\pm4$&$(3.4 \pm 0.7)\times 10^{24}$\\
HCN&1&4&3&354.505&42.54&$1540\pm31$&$(3.0 \pm 0.1) \times 10^{26}$\\
CH$_3$OH&2&$4_{0}$&$3_{-1}$ E&350.688&36.34&$36\pm4$&$(5.8 \pm0.9)\times10^{27}$\\
CH$_3$OH&2&$1_{1}$&$0_{0}$ A+&350.905&16.84&$50\pm4$&"\\
CH$_3$OH&2&$7_{2}$&$6_{1}$ E&363.740&87.26&$73\pm6$&"\\
\hline
\end{tabular}
}
\end{table*}

\begin{figure}
\labellist
\pinlabel a) at 10 440
\endlabellist
\centering
\hspace{-1mm}
\includegraphics[width=\columnwidth]{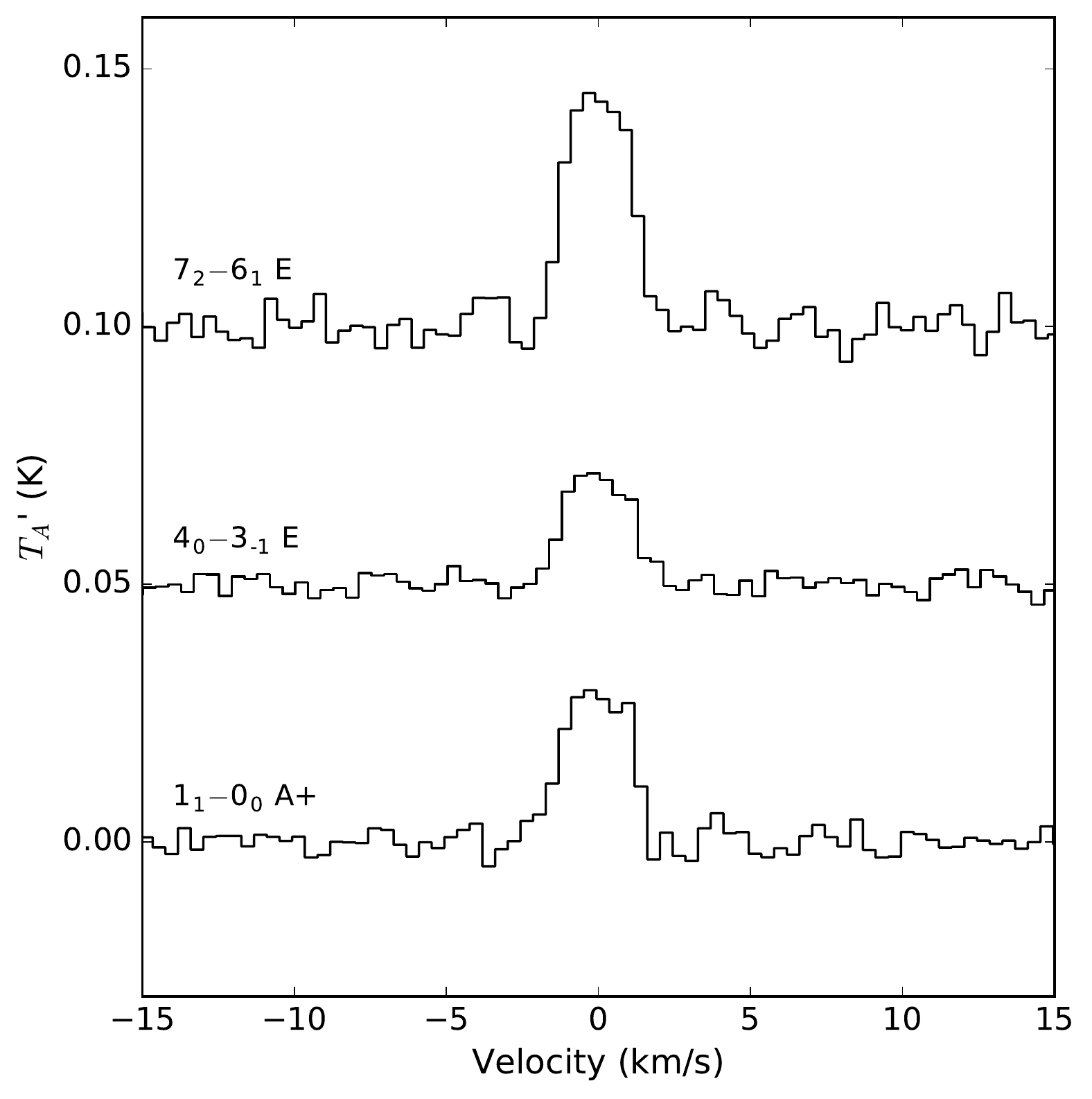}\\[2mm]
\labellist
\pinlabel b) at 10 405
\endlabellist
\includegraphics[width=\columnwidth]{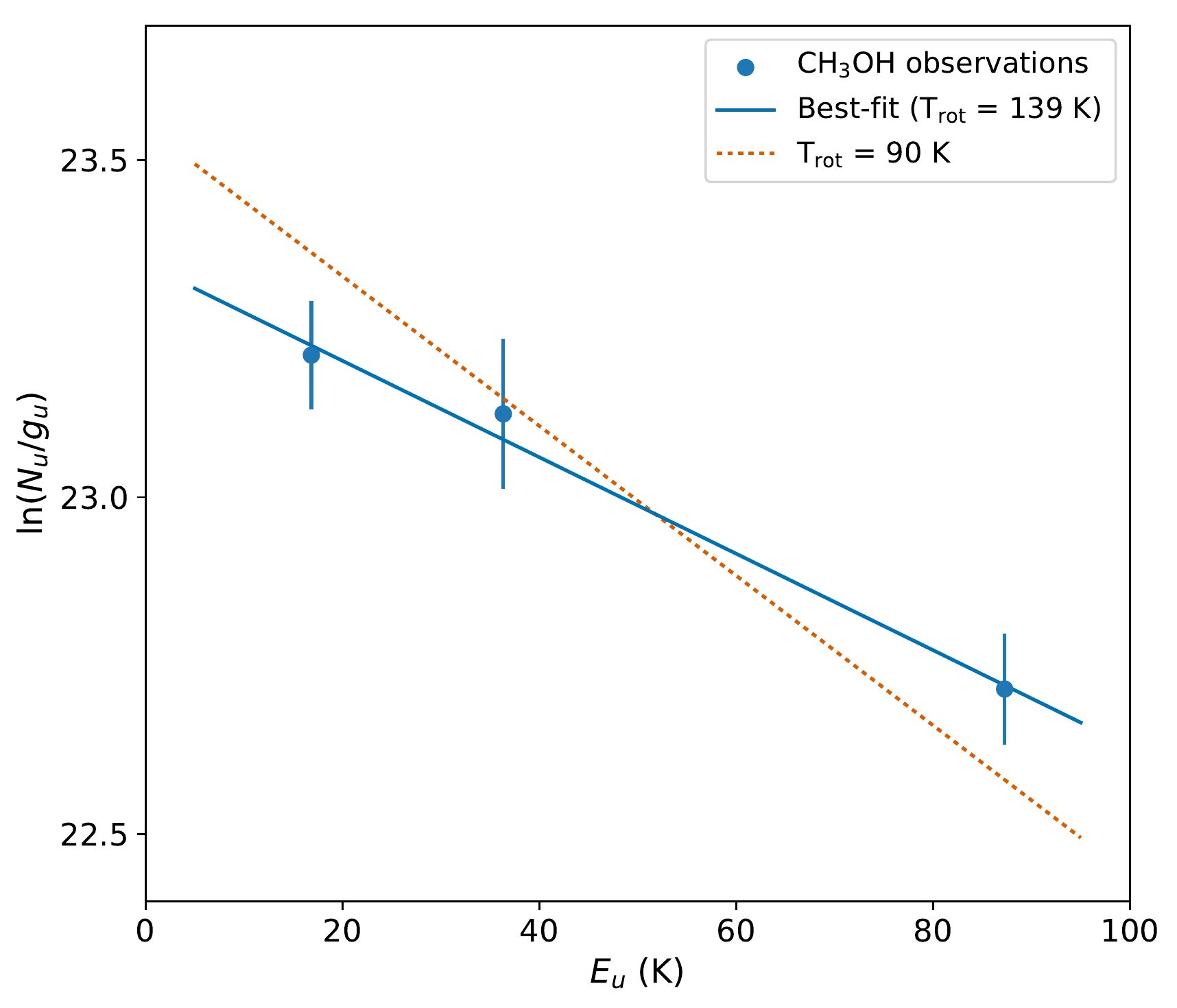}
\caption{(a) The three observed CH$_3$OH lines, offset vertically for clarity.  (b) CH$_3$OH rotational diagram (filled blue circles) and (solid) line of best fit, with $T_{rot}=139$~K. The $T_{rot}=90$~K (dotted red) line determined by \citet{agu14} on 2013 November 14 is shown for reference.}
\label{fig:rotdiag}
\end{figure}

Three CH$_3$OH lines with differing upper-state energies were measured as a probe of the coma temperature (shown in Figure \ref{fig:rotdiag}a). Using a rotational diagram analysis \citep[\emph{e.g.}][]{boc94}, the best-fitting rotational temperature was found to be $T_{rot}=139\pm29$~K (Figure \ref{fig:rotdiag}b).

\section{Radiative transfer modeling}
    
Local thermodynamic equilibrium (LTE) breaks down at distances greater than $\sim1000$~km ($1.6''$) from the comet, so to derive production rates we solve the molecular excitation and radiation transfer in the coma. Our model employs a modified version of the Line Modeling Engine \citep[LIME;][]{bri10}, following a similar approach to \citet{dev18} and \citet{bog17}, combining a statistical equilibrium calculation with Monte Carlo photon propagation. We also include collisions with electrons and fluorescent pumping by Solar radiation, which become important on the large coma scales probed by our autocorrelation observations. The model coma is divided into 5,000 cells on an unstructured Delaunay grid with logarithmic radial sampling. {We adopted a standard}, spherically-symmetric \citet{has57} density distribution, with constant kinetic temperature and expansion velocity of 1.1~km\,s$^{-1}$ (determined from the HCN line width).  A water production rate $Q({\rm H_2O}) = 2\times10^{29}$~s$^{-1}$ was used, following \citet{cor14}.

Molecular collision rates for HCN and CH$_3$OH were taken from the Leiden Atomic and Molecular Database \citep[LAMDA;][]{sch05}, scaling H$_2$ collisional rates by the square root of the reduced mass ratio, to approximate collisions with H$_2$O (the dominant coma gas). Electron collisional rates were calculated following \citet{zak07}. We investigated the importance of uncertainties in the collisional rates by varying them by a factor of 2 and found that the predicted line intensities changed by only a few percent. Hyperfine structure was modeled for HCN assuming equilibrium line strength ratios for the components in a given $J$ level. Photodissociation rates for HCN and CH$_3$OH were obtained from \cite{cro94}. Fluorescent pumping was calculated using the method of \citet{cro83}, and the photodissociation and pumping rates for H$^{13}$CN were assumed to be the same as for HCN. Model image cubes were convolved with the measured ALMA 12-m primary beam pattern. Finally, model spectra were extracted at the nucleus position identified by \citep{cor14}, for comparison with observations.

The coma kinetic temperature was derived by simultaneously fitting all the detected CH$_3$OH transitions, resulting in $T_{kin}= 145$~K. This is only slightly larger than the rotational temperature $T_{rot}= 139$~K, indicating the observed CH$_3$OH was close to LTE.  Using $T_{kin}= 145$~K, we derived production rates ($Q$) for the observed species, given in Table \ref{tab:lines}.  

\section{Discussion}

Our ALMA observations and modeling of comet ISON reveal an \ratio\ ratio of $88 \pm 18$, which is consistent with the terrestrial \ratioc\ value of 89.4 \citep{cop02}, and suggests a lack of strong $^{13}$C fractionation in HCN during comet formation in the protosolar disk. {This} result is also consistent with the error-weighted mean ratio of $111\pm9$ observed in previous comets: C/1995 O1 (Hale-Bopp; \citealt{jew97}), 17P/Holmes \citep{boc08}, C/2014 Q2 (Lovejoy) and C/2012 F6 (Lemmon) \citep{biv16}. However, the relatively low \ratio\ value in C/2012 S1 (ISON) tends to weaken \citet{biv16}'s claim of marginal evidence for $^{13}$C depletion in this molecule. Thus, accounting for errors, the average \ratio\ ratio in the overall comet population may still be consistent with the value $\approx90$ found in cometary CN and C$_2$ \citep{boc15}, and in carbonaceous materials from a range of Solar System bodies \citep{woo09,woo09b}. Although our measured \ratio\ ratio does not completely rule out the survival of some $^{13}$C-depleted HCN for incorporation into comets (for example, from a cold interstellar cloud or protoplanetary disk phase), it implies that such fractionation effects {were} relatively small.

Our CH$_3$OH excitation analysis indicates a rotational temperature of $T_{rot}=139$~K, which is significantly larger than the value of 90~K found three days earlier by \citet{agu14} using the IRAM 30-m telescope (with a $10''$ beam). Their analysis incorporated many more CH$_3$OH lines (in the 249-267~GHz range), and should therefore be reliable. However, the lower value of 90~K is incompatible with our CH$_3$OH observations (see Figure \ref{fig:rotdiag}b). A sharp increase in coma temperature between November 14-17 could be explained by an increase in the coma heat input (or heating efficiency). Coma temperatures have been theorized to correlate with the water production rate \citep{boc87,com04}. \citet{agu14} and \citet{cor17} both identified strong coma variability during November 13-16, which could therefore have been associated with corresponding temperature fluctuations. Coma heating by the sublimation of icy grains released episodically from the nucleus around the time of our observations could also have helped raise the coma temperature \citep[\emph{e.g.}][]{fou12}. However, given that we only observed three CH$_3$OH lines, an excitation anomaly or other issue resulting in spurious CH$_3$OH line strengths cannot be entirely ruled out. Thus, we ran alternative models using $T_{rot}=90$~K ($T_{kin}=100$~K) and found $Q({\rm HCN})=2.9\times10^{26}$~s$^{-1}$, and $Q({\rm H^{13}CN})=3.5\times10^{24}$~s$^{-1}$. The corresponding \ratio\ ratio is $82\pm17$, which reinforces our overall conclusion regarding the lack of significant $^{13}$C fractionation.

{Our HCN production rate of $(3.0 \pm 0.1) \times 10^{26}$~s$^{-1}$ is reasonably close to the values of $3.5\times10^{26}$~s$^{-1}$ and $(4.0\pm0.5)\times10^{26}$~s$^{-1}$ obtained by \citet{cor14} and \citet{bog17}, respectively, using well-calibrated interferometric data from the same date. Discrepancies between these $Q({\rm HCN})$ values could be explained by differences in modeling approach, as well as differences in the spatial sampling of the interferometric and autocorrelation data, for example, if the HCN temperature, outflow velocity or production rate varied as a function of coma radius.

The extremely low (1.4~mK) RMS noise level of our H$^{13}$CN spectrum, with only 45 min of observations, demonstrates the great potential of ALMA autocorrelation spectroscopy for detecting weak spectral lines in extended sources. Such observations could, in the future, allow for more efficient use of observatory resources to detect new molecules. While standalone autocorrelation observations are not currently offered by the observatory, using the four antennas of the ALMA total power (TP) array, it would take 3.5 times as long to reach the same noise level provided by the $50\times$12-m antennas of the main array. The ACA ($12\times$7~m array) has a synthesized beam FWHM about 4.6 times smaller than the primary beam of the (12~m) total power antennas. Consequently, for very extended sources (such as comets and nearby interstellar clouds), the ACA gathers less flux per beam and therefore would require significantly longer to reach the same extended-source sensitivity as the dedicated ($4\times12$ m) TP array, despite having a similar collecting area.}

{Our ad-hoc use of phase calibration scans to subtract the sky and telescope background signal is far from ideal, as highlighted by the inability to fully cancel standing waves and strong sky emission lines. The statistical noise is also unnecessarily high due to the short duration of our off-source (reference) scans relative to the on-source time. Both these factors adversely impacted the quality of our spectra. For ALMA autocorrelation capabilities to be fully realized, it will be necessary to implement a dedicated position-switching mode for the main array, allowing full control over the source and reference durations and positions, in order to achieve optimal background cancellation and further improve the sensitivity to weak spectral lines.}

\section{Conclusion}

We have demonstrated the first scientific use of ALMA autocorrelation spectra from the main 12-m array to observe extended sky emission that was resolved out by the interferometer. This technique has been applied to archival observations of comet C/2012 S1 (ISON). The resulting HCN spectra are several times more sensitive to coma emission than the spatially-integrated interferometric data, allowing a significant improvement in our ability to detect weak spectral lines. Thus, we obtained the first detection of the rare isotopologue H$^{13}$CN in this comet. Using a newly developed non-LTE radiative transfer model including collisions with H$_2$O, electrons and fluorescent pumping, we derived a \ratioc\ ratio of $88 \pm 18$. This result is consistent with the value of $\approx90$ found in carbonaceous material from a range of Solar System bodies, and, combined with prior cometary \ratioc\ measurements, implies a lack of $^{13}$C fractionation during the formation of cometary organics in the protosolar accretion disk or prior interstellar cloud.

These results demonstrate the potential for significant improvements in sensitivity for spectroscopic observations of cometary comae and other extended sources using autocorrelation data from the main array of {(up to 50)} 12-m ALMA antennas. This technique can be usefully applied to help retrieve extended flux from any archival ALMA dataset (in particular, those lacking dedicated total power observations), and could enable a significant expansion of ALMA's scientific return {in searches for weak spectral lines in extended objects}, especially for observations of transient targets, for which simultaneous single-dish observations may not be possible.

\acknowledgments
This work was supported by the National Science Foundation (under Grant No. AST-1614471) and the NASA Astrobiology Institute through the Goddard Center for Astrobiology. It makes use of ALMA data sets ADS/JAO.ALMA\#2012.A.00033.S and 2013.1.00278.S. ALMA is a partnership of ESO, NSF (USA), NINS (Japan), NRC (Canada), NSC and ASIAA (Taiwan) and KASI (Republic of Korea), in cooperation with the Republic of Chile. The JAO is operated by ESO, AUI/NRAO and NAOJ.  The NRAO is a facility of the National Science Foundation operated under cooperative agreement by Associated Universities, Inc.


\end{document}